\documentclass[prd,twocolumn,floatfix,preprintnumbers,showpacs]{revtex4}
\usepackage{graphics}
\usepackage{epsfig}

\begin{document}
\title{Probing the dark side:
Constraints on the dark energy equation of state
from CMB, large scale structure and Type Ia supernovae}
\author{Steen Hannestad}
\email{steen@nordita.dk}
\affiliation{NORDITA, Blegdamsvej 17, DK-2100 Copenhagen, Denmark}
\author{Edvard M{\"o}rtsell}
\email{edvard@physto.se}
\affiliation{Department of Physics, Stockholm University, S - 106 91 Stockholm, Sweden}
\date{{\today}}

\preprint{NSF-ITP-02-13}

\begin{abstract}
  We have reanalysed constraints on the equation of state parameter,
  $w_Q\equiv P/\rho$, of the dark energy, using several cosmological
  data sets and relaxing the usual constraint $w_Q \geq -1$. We find
  that combining Cosmic Microwave Background, large scale structure
  and Type Ia supernova data yields a non-trivial lower bound on
  $w_Q$.  At 95.4\,\% confidence we find, assuming a flat geometry of
  the universe, a bound of $-2.68 < w_Q < -0.78$ if $w_Q$~is taken to
  be a completely free parameter.  Reassuringly we also find that the
  constraint $w_Q \geq -1$ does not significantly bias the overall
  allowed region for $w_Q$. When this constraint is imposed the
  95.4\,\% confidence bound is $-1 \leq w_Q < -0.71$. Also, a pure
  cosmological constant $(w = -1)$ is an excellent fit to all
  available data.  Based on simulations of future data from the Planck
  CMB experiment and the SNAP and SNfactory supernova experiments we
  estimate that it will be possible to constrain $w_Q$ at the 5\,\%
  level in the future.

\end{abstract}
\pacs{98.80.-k,12.60.-i,14.80.-j} 
\maketitle

%%%%%%%%%%%%%%%%%%%%%%%%%%%%%%%%%%%%%%%%%%%%%%%%%%%%%%%%%%%%%%%%%%%%%%
%%%% Section I %%%%%%%%%%%%%%%%%%%%%%%%%%%%%%%%%%%%%%%%%%%%%%%%%%%%%%%
%%%%%%%%%%%%%%%%%%%%%%%%%%%%%%%%%%%%%%%%%%%%%%%%%%%%%%%%%%%%%%%%%%%%%%

\section{introduction}

Several independent methods of observation all suggest that most of
the energy density in the universe is in the form of a component with
negative pressure -- dark energy. The simplest possibility for such a
component is a cosmological constant which has a constant equation of
state $P_Q = - \rho_Q$. However, in the general case the dark energy
can have an equation of state which is time-dependent, $P_Q = w_Q(t)
\rho_Q$ \cite{Wetterich:fm,Peebles:1987ek,Wetterich:bg,Zlatev:1998tr,
  Wang:1999fa,Steinhardt:nw,Wang:2001ht,Wang:2001da,Baccigalupi:2001aa}. 
Dark energy with a time-dependent
equation of state has been invoked to explain the coincidence problem,
the fact that the energy density in dark energy is roughly equal to
that in dark matter exactly at the present epoch. By coupling a scalar
field to matter one can obtain tracking solutions for the time
dependence of the dark energy density so that it always follows the
dominant energy density component.

Generic to most of these proposed candidates for dark energy is the
fact that $w_Q \geq -1$ at all times. This is, e.g., the case for most
quintessence models where a scalar field is rolling in a potential
(potential driven quintessence). Since most of the plausible models
lie in this category, the likelihood analyses which have been used to
find $w_Q$ have cut away the region with $w_Q<-1$. From a purely
phenomenological point of view this is not justified and can lead to
severe bias in parameter determination.  This is particularly
worrisome because the most recent data set an upper limit on
$w_Q<-0.85$ (68\,\% confidence) \cite{Bean:2001xy}
(see also \cite{Corasaniti:2001mf}), but {\it not} a
lower limit.  Therefore, if the true model has $w_Q<-1$, the upper
bound could be wrong (and in principle a full analysis could even rule
out a cosmological constant as being the dark energy). In fact there
are several models which predict a dark energy component with $w_Q<-1$
\cite{Caldwell:1999ew,Chiba:1999ka,Ziaeepour:2000qq,Faraoni:2001tq,Riazuelo:2000fc,Schulz:2001yx}. This possible bias problem was also discussed in
Ref.~\cite{Maor:2001ku}.

In the present paper we reanalyse cosmological data from the Cosmic
Microwave Background (CMB), large scale structure (LSS) and Type Ia
supernovae (SNe) without the constraint $w_Q \geq -1$. We make the
simplifying approximation that $w_Q(t) =$ const. Even though this is
certainly not true for many models of dark energy, almost all models
can be very well approximated by a model having a constant $w_{Q,{\rm
    eff}}$, which is then calculated as a properly weighted mean of
$w_Q(t)$ \cite{Huey:1998se,Bean:2001xy}.  With this approximation the
dark energy density evolves simply as $\rho_Q \propto a^{-3(1+w_Q)}$,
where $a$ is the scale factor.

Our analysis of the present data is in many ways similar to that
performed in Ref.~\cite{Bean:2001xy}. However, in addition to the
extension of the parameter space to $w_Q < -1$ we also include new
data from the 2dF galaxy survey.

Finally, we discuss the prospects for measuring $w_Q$ precisely with
future high precision CMB and SN data.

%%%%%%%%%%%%%%%%%%%%%%%%%%%%%%%%%%%%%%%%%%%%%%%%%%%%%%%%%%%%%%%%%%%%%%
%%%% Section II %%%%%%%%%%%%%%%%%%%%%%%%%%%%%%%%%%%%%%%%%%%%%%%%%%%%%%
%%%%%%%%%%%%%%%%%%%%%%%%%%%%%%%%%%%%%%%%%%%%%%%%%%%%%%%%%%%%%%%%%%%%%%

\section{Data analysis}

\subsection{CMB and large scale structure}

{\it CMB data set ---} Several data sets of high precision are now
publicly available.  In addition to the COBE \cite{Bennett:1996ce}
data for small $l$ there are data from BOOMERANG \cite{boom}, MAXIMA
\cite{max}, DASI \cite{dasi} and several other experiments
\cite{WTZ,qmask}.  Wang, Tegmark and Zaldarriaga \cite{WTZ} have
compiled a combined data set from all these available data, including
calibration errors.  In the present work we use this compiled data
set, which is both easy to use and includes all relevant present
information.  

{\it LSS data set ---} At present, by far the largest survey available
is the 2dF \cite{2dF} of which about 147,000 galaxies have so far been
analysed. Tegmark, Hamilton and Xu \cite{THX} have calculated a power
spectrum, $P(k)$, from this data, which we use in the present work.
The 2dF data extends to very small scales where there are large
effects of non-linearity. Since we only calculate linear power
spectra, we use (in accordance with standard procedure) only data on
scales larger than $k = 0.2 h \,\, {\rm Mpc}^{-1}$, where effects of
non-linearity should be minimal.

The CMB fluctuations are usually described in terms of the power
spectrum, which is again expressed in terms of $C_l$ coefficients as
$l(l+1)C_l$, where
\begin{equation}
C_l \equiv \langle |a_{lm}|^2\rangle.
\end{equation}
The $a_{lm}$ coefficients are given in terms of the actual temperature
fluctuations as
\begin{equation}
T(\theta,\phi) = \sum_{lm} a_{lm} Y_{lm} (\theta,\phi).
\end{equation}
Given a set of experimental measurements, the likelihood function is
\begin{equation}
{\cal L}(\Theta) \propto \exp \left( -\frac{1}{2} x^\dagger
[C(\Theta)^{-1}] x \right),
\end{equation}
where $\Theta = (\Omega, \Omega_b, H_0, n, \tau, \ldots)$ is a vector
describing the given point in parameter space, $x$ is a vector
containing all the data points, and $C(\Theta)$ is the data covariance
matrix.  This applies when the errors are Gaussian. If we also assume
that the errors are uncorrelated, it can be reduced to the simple
expression, ${\cal L} \propto e^{-\chi^2/2}$, where
\begin{equation}
\chi^2 = \sum_{i=1}^{N_{\rm max}} \frac{(C_{l, {\rm obs}}-C_{l,{\rm
theory}})_i^2} {\sigma(C_l)_i^2},
\label{eq:chi2}
\end{equation} 
is a $\chi^2$-statistic and $N_{\rm max}$ is the number of power
spectrum data points \cite{oh}.  In the present paper we use
Eq.~(\ref{eq:chi2}) for calculating $\chi^2$ for the CMB data.  Since
we also use data from the 2dF survey the total $\chi^2$ is then given
by

\begin{eqnarray}
\chi^2 &=&  \sum_{i=1}^{N_{\rm max,CMB}} \frac{(C_{l, {\rm obs}}-C_{l,{\rm
theory}})_i^2} {\sigma(C_l)_i^2}\nonumber\\
&+& \sum_{j=1}^{N_{\rm max,LSS}} \frac{(P(k)_{{\rm obs}}-P(k)_{{\rm
theory}})_j^2} {\sigma(P(k))_j^2}.
\label{eq:chi22}
\end{eqnarray} 

The procedure is then to calculate the likelihood function over the
space of cosmological parameters. For calculating CMB and matter
power spectra we have used the publicly available CMBFAST package
\cite{CMBFAST}.
The 2D likelihood function for $(\Omega_m,w_Q)$ is obtained by keeping
$(\Omega_m,w_Q)$ fixed and marginalising over all other parameters.

As free parameters in the likelihood analysis we use $\Omega_m$, the
matter density, $w_Q$, the dark energy equation of state, $\Omega_b$,
the baryon density, $H_0$, the Hubble parameter, $n$, the scalar
spectral index, and $\tau$, the optical depth to reionization.  The
normalisation of the CMB data, $Q$, and of the 2dF data, $b$, are
taken as completely free and uncorrelated parameters in the analysis.
This is very conservative and eliminates any possible systematics
involved in determining the bias parameter.  We constrain the analysis
to flat ($\Omega_m + \Omega_Q = 1$) models, and we assume that the
tensor mode contribution is negligible ($T/S=0$).  These assumptions
are compatible with analyses of the present data \cite{WTZ}, and
relaxing them does not have a big effect on the final results.

\begin{center}
\begin{table}[hb]
\caption{The different priors used in the analysis.}
\begin{tabular}{lc}
\colrule
Parameter & Prior \cr
%\colrule 
\colrule
$\Omega_m$ & 0.1-1 \cr
$w_Q$ & -3 \,\, - \,\, -0.5 \cr
$\Omega_b h^2$ & $0.020 \pm 0.002$ (Gaussian) \cr
$h$ & $0.72 \pm 0.08$ (Gaussian) \cr
$n$ & 0.5-1.4 \cr
$\tau$ & 0-1 \cr
$Q$ & free \cr 
$b$ & free \cr
\colrule
\end{tabular}
\end{table}
\end{center}

Table I shows the different priors used. We use the constraint $H_0 =
72 \pm 8 \,\, {\rm km} \, {\rm s}^{-1} \, {\rm Mpc}^{-1}$ [$h \equiv
H_0/(100 \,\, {\rm km} \, {\rm s}^{-1} \, {\rm Mpc}^{-1})$] from the
HST Hubble key project \cite{freedman} (the constraint is added
assuming a Gaussian distribution) and the constraint $\Omega_b h^2 =
0.020 \pm 0.002$ from BBN \cite{Burles:2000zk}.

Fig.~1 then shows the 68.3\,\% and 95.4\,\% confidence allowed
regions, corresponding to $\Delta \chi^2 = 2.31$ and 6.17
respectively.

% --------------------------------------
% Figure 1
% --------------------------------------
\begin{figure}[h]
\begin{center}
\epsfysize=7truecm\epsfbox{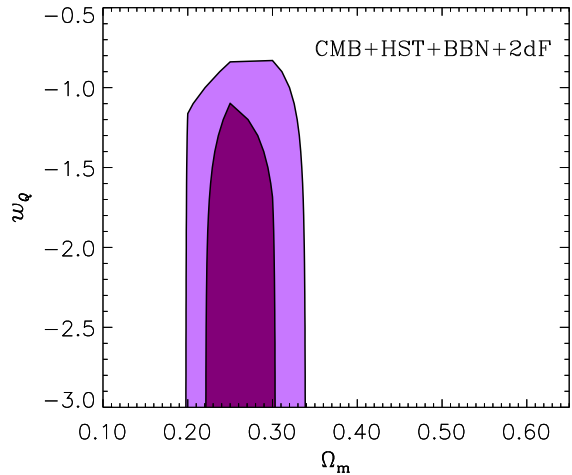}
\vspace{0truecm}
\end{center}
\caption{The 68.3\,\% (dark shaded) and 95.4\,\% (light shaded) 
confidence allowed regions for $\Omega_m$ and
$w_Q$ using CMB, HST, BBN and LSS data.}
\label{fig1}
\end{figure}

For very negative values of $w_Q$ the CMB and LSS constraint is
independent of $w_Q$. The reason is that at such low $w_Q$ the dark
energy influences the CMB spectrum only via the late integrated
Sachs-Wolfe effect. However, the late ISW effect decreases in
magnitude as $w_Q$ decreases. This effect can be seen in Fig.~2 where
CMB power spectra are plotted for different values of $w_Q$.

% --------------------------------------
% Figure 2
% --------------------------------------
\begin{figure}[h]
\begin{center}
\epsfysize=7truecm\epsfbox{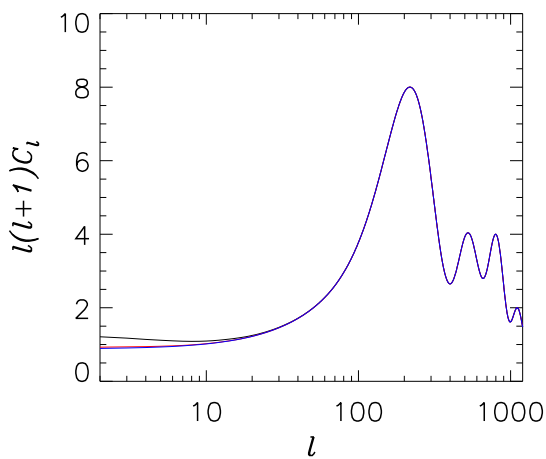}
\vspace{0truecm}
\end{center}
\caption{Different CMB power spectra for different values of 
$w_Q$. In all cases the model parameters are those of the fiducial
$\Lambda$CDM model, $\Omega_m=0.3$, $\Omega_\Lambda=0.7$, 
$\Omega_b h^2 = 0.020$, $H_0 = 70 \,\, {\rm km} \, {\rm s}^{-1} \, 
{\rm Mpc}^{-1}$. The full line is for $w_Q=-1$, the dashed for $w_Q=-2$,
and the dotted for $w_Q=-4$ (in order of decreasing $C_l$ at low $l$).}
\label{fig2}
\end{figure}

\subsection{Type Ia supernovae}
{\it SN data set ---} The SN data set used in this analysis
corresponds to Fit~C from the Supernova Cosmology Project as described
in \cite{perlmutter}. This is a subsample of a total of 60 SNe where
two SNe are excluded as statistical outliers, two because of atypical
lightcurves and two because of suspected reddening.

We use this data set to fit $\Omega_m$ and $w_Q$, taking advantage of
the cosmology dependence of the distance-redshift relation.  Type~Ia
SNe are very useful as distance indicators because of their high
luminosities and small dispersion among their peak absolute magnitudes
($\sigma_m\simeq$ 0.15).  Also, they have distinct spectral lines,
allowing for accurate redshift determinations.

The apparent and absolute magnitudes are related by
\begin{equation}\label{eq:mz} 
  m(z)=M+5\log\left[{\cal D}_{L}(z,\Omega_m,w_Q)\right]-5\log H_0+25,
\end{equation}
where ${\cal D}_{L}:=H_0d_{L}$ is the part of the luminosity distance
that remains after multiplying out the dependence on the Hubble
constant (expressed here in units of km s$^{-1}$ Mpc$^{-1}$). In the
low redshift limit, Eq.~(\ref{eq:mz}) reduces to a linear Hubble
relation between $m$ and $\log z$:
\begin{equation}
  m(z)={\cal M}+5\log z,
\end{equation}
where we have expressed the intercept of the Hubble line as ${\cal
M}:=M-5\log H_0+25$. This quantity can be measured from the apparent
magnitude of low redshift standard candles, without knowing the value
of $H_0$.  Thus, with a set of apparent magnitude and redshift
measurements $m(z)$ for Type Ia SNe, we can find the best-fit values
of $(\Omega_m, w_Q)$ to solve the equation
\begin{equation}
  m(z)-{\cal M}=5\log\left[{\cal D}_{L}(z,\Omega_m,w_Q)\right].
\end{equation}

The $\chi^2$ is then given by
\begin{equation}
  \chi^2=\sum_{i=1}^n\frac{\left[m_i-5\log\left[{\cal D}_{L}(z_i,\Omega_m,w_Q)\right] -{\cal
  M}\right]^2}{\sigma_i^2},
\end{equation}
where $\sigma_i$ is the statistical uncertainty for each event, We
assume a flat geometry of the universe when calculating ${\cal D}_{L}$
and marginalise over ${\cal M}$ where we assume no prior knowledge.

\begin{figure}[h]
\begin{center}
\epsfysize=7truecm\epsfbox{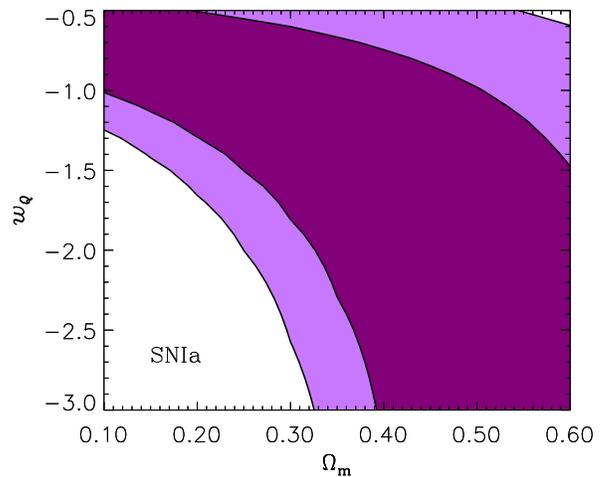}
\vspace{0truecm}
\end{center}
\caption{The 68.3\,\% (dark shaded) and 95.4\,\% (light shaded) 
confidence allowed regions for $\Omega_m$ and $w_Q$ using the 54 type
Ia SNe from the Supernova Cosmology Project.}
\label{fig3}
\end{figure}

In Fig.~\ref{fig3}, the 68.3\,\% and 95.4\,\% confidence allowed
regions are showed. The best-fit values are $\Omega_m=0.45$ and
$w_Q=-1.9$, indicating the possibility of a bias in the parameter
determination when imposing the constraint $w_Q \geq -1$.

% --------------------------------------
% Figure 3
% --------------------------------------

\subsection{Combined constraint}

When all the available data is combined, a fairly stringent bound on
$w_Q$~is obtained. The 68.3\,\% and 95.4\,\% confidence combined
bounds are shown in Fig.~4. For $w_Q$~alone we find a 95.4\,\%
confidence bound of $-2.68 < w_Q < -0.78$.

% --------------------------------------
% Figure 4
% --------------------------------------
\begin{figure}[h]
\begin{center}
\epsfysize=7truecm\epsfbox{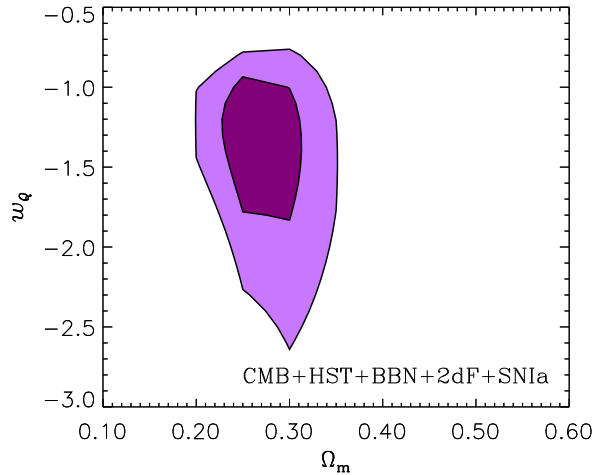}
\vspace{0truecm}
\end{center}
\caption{The 68.3\,\% (dark shaded) and 95.4\,\% (light shaded) 
confidence allowed regions for $\Omega_m$ and
$w_Q$ using all available data.}
\label{fig4}
\end{figure}

\section{discussion}

\subsection{Constraints from present data} 

An important point is to determine whether the relaxation of the bound
$w_Q\geq -1$ significantly affects the likelihood analysis for the
part of parameter space which is above $w_Q=-1$. This could be the
case if the best-fit value of $w_Q$ lies in the excluded region as is
the case for the current SN data, where the best-fit value corresponds
to $\Omega_m=0.45$, $w_Q=-1.9$. For the CMB+LSS data the situation is
the same, with the overall best fit being at $\Omega_m=0.26$,
$w_Q=-2.6$.  In Fig.~5 we plot the the likelihood contours with the
constraint $w_Q \geq -1$ imposed. In terms of $w_Q$~alone the 95.4\,\%
confidence bound is now $-1 \leq w_Q < -0.71$.

% --------------------------------------
% Figure 5
% --------------------------------------
\begin{figure}[h]
\begin{center}
\epsfysize=7truecm\epsfbox{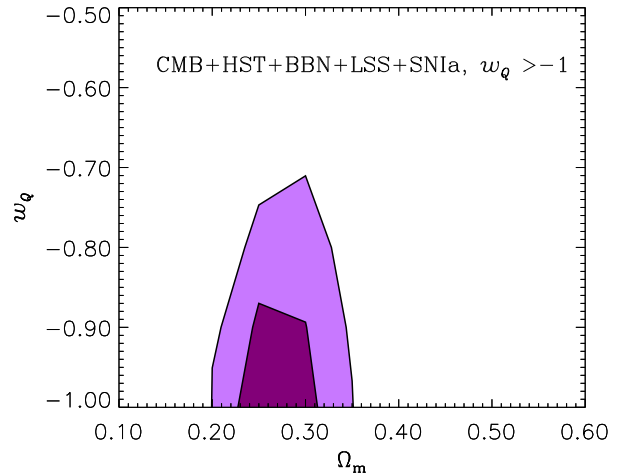}
\vspace{0truecm}
\end{center}
\caption{The 68.3\,\% (dark shaded) and 95.4\,\% (light shaded) 
confidence allowed regions for $\Omega_m$ and $w_Q$ using all
available data and imposing the bound $w_Q \geq -1$.}
\label{fig5}
\end{figure}

Comparing this to the bound obtained without the constraint shows that
the likelihood contours are fairly similar and that the bias is not a
significant problem. Our Fig.~4 can for instance also be compared to
Fig.~3 of Ref.~\cite{Bean:2001xy}.  Apart from the smaller allowed
region in $\Omega_m$ because of the tighter BBN constraint and the use
of the full 2dF data set, the two plots are very similar. The bound on
$w_Q$ found in Ref.~\cite{Bean:2001xy} is $-1 \leq w < -0.73$, very
similar to the constraint found in the present analysis when the
constraint $w_Q \geq -1$ is imposed.

This shows both that our constrained analysis is consistent with
Ref.~\cite{Bean:2001xy} and that this previous
analysis of $w_Q$~is not seriously biased.

The next very important point of our analysis is that one obtains a
non-trivial lower bound on $w_Q$ from the combination of CMB, LSS and
SN data. From SN data alone we can infer that $w_Q \gtrsim -12$ is
ruled out at the 68.3\,\% confidence level, but combining this with
CMB and LSS data tightens the bound significantly to $w_Q > -2.68$ at
95.4\,\% confidence. It is also very interesting and perhaps somewhat
suggestive that a cosmological constant lies in the 68.3\,\%
confidence allowed region.

\subsection{Constraints from future data}
The ability to constrain the equation of state parameter $w_Q$ of a
dark energy component using future CMB and Type Ia SN data have been
recently investigated by a number of authors, see, e.g.,
\cite{goliath} and references therein. Our analysis differ in the
respect that we do not impose the constraint $w_Q\geq -1$ and that we
use the most current anticipated data sets.

{\it CMB data set ---}For CMB we use simulated data from the Planck
Surveyor satellite. For simplicity we use only data from the HFI 100
GHz channel, assuming an angular resolution of 10.7 arcmin and a pixel
noise of $\Delta T/T = 1.7 \times 10^{-6}$ \cite{Planck}. This channel
is not polarization sensitive and so our assumed data set seems
conservative compared to what can be expected from the full Planck
data. On the other hand we do not include foregrounds in our analysis.
The simulated data is generated from an underlying flat model with the
following parameters: $\Omega_m = 0.3$, $w_Q = -1$, $\Omega_b h^2 =
0.02$, $H_0 = 70 \,\,{\rm km} \, {\rm s}^{-1} \, {\rm Mpc}^{-1}$, $n =
1.0$, and $\tau = 0$.

{\it SN data set ---} We use simulated data sets corresponding to
three year's data from the proposed satellite telescope the
Supernova/Acceleration Probe (SNAP; \cite{snap}) and the predicted
results from the Supernova Factory Campaign (SNfactory; \cite{snfac})
scheduled to start in the end of 2002.

The SNAP satellite would be capable of discovering and taking spectra
of $\sim 2800$ Type SNe per year for redshifts $z<1.7$. The current
projected redshift distribution follows the distribution in the SNAP
proposal \cite{snap} for $z<1$ and is approximately uniform at higher
redshifts \cite{linder}. The SNfactory data set consists of 200~SNe
between $0.03<z<0.06$ and 100~SNe between $0.06<z<0.15$.  The
simulated data is generated for the same underlying model as for the
CMB simulations.

% --------------------------------------
% Figure 6
% --------------------------------------
\begin{figure}[h]
\begin{center}
\epsfysize=12truecm\epsfbox{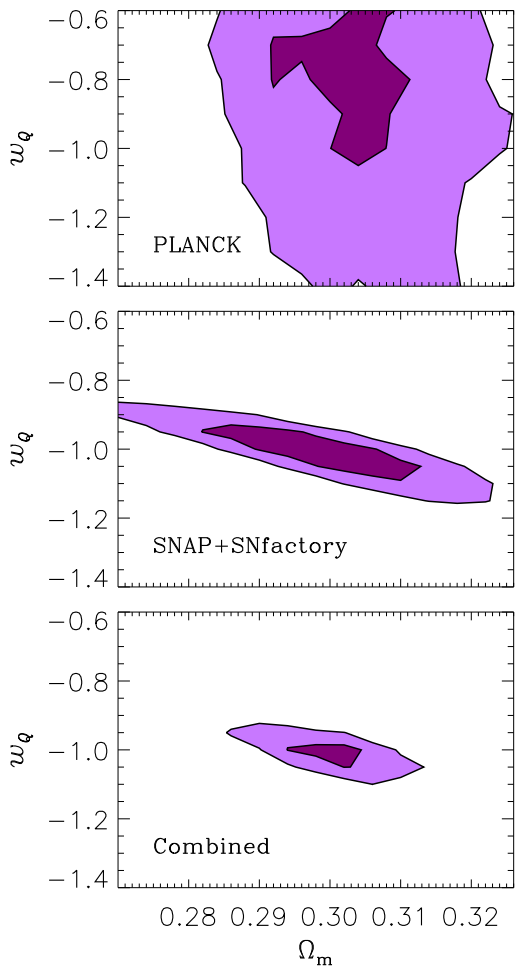}
\vspace{0.5truecm}
\end{center}
\caption{The 68.3\,\% (dark shaded) and 95.4\,\% (light shaded) 
  confidence allowed regions for $\Omega_m$ and $w_Q$ using simulated
  data from the Planck satellite and the SNAP and SNfactory
  observatories.}
\label{fig6}
\end{figure}

In Fig.~6 we show a likelihood analysis based on the simulated CMB and
Type Ia SN data. It is clear that again the data will complement each
other, the CMB data being very sensitive to $\Omega_m$ and the SNIa
data to $w_Q$. From the combined data set we estimate that it is
possible to obtain a 95.4\% confidence interval on $w_Q$ of roughly
0.05 relative precision in the two parameters. In this note we have
neglected the possible use of multiple imaged core-collapse SNe to
constrain $w_Q$ and $\Omega_m$, see Ref.~\cite{multisne}.

It is important to note that the confidence regions presented only
take into account statistical uncertainties. For future Type Ia SN
data sets, systematic errors from, e.g., dust obscuration, luminosity
evolution and gravitational lensing might be comparable to or even
larger than the statistical errors. In Fig.~7 the confidence regions
corresponding to the middle panel of Fig.~6 is shown if lensing
effects from 90\,\% NFW dark matter halos and 10\,\% point masses are
included in the simulated data set (generated with $\Omega_m = 0.3$
and $w_Q = -1$). It is obvious that gravitational lensing, if not
taken into account, will cause an underestimation of the matter
density and in order to reach the full potential of the large
statistics, we need to to correct for the effect, e.g., as described
in \cite{amanullah}. For the CMB data there may also be systematic
errors of a magnitude comparable to the purely statistical ones.
 
However, if the systematic errors can be controlled in an effective
matter, we conclude that it should be possible to constrain the
equation of state parameter $w_Q$ of the dark energy to high accuracy
using future CMB and SN data sets.

% --------------------------------------
% Figure 7
% --------------------------------------
\begin{figure}[h]
\begin{center}
\epsfysize=6truecm\epsfbox{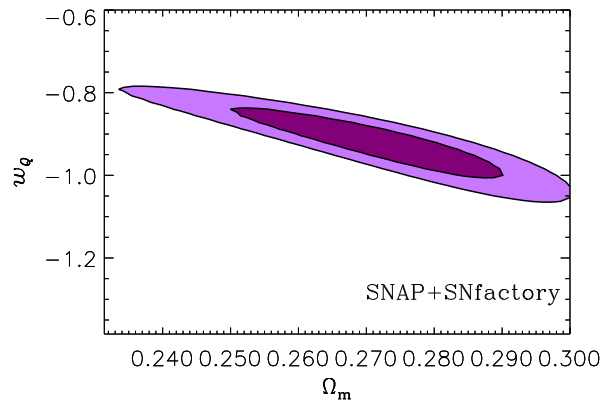}
\end{center}
\caption{The 68.3\,\% (dark shaded) and 95.4\,\% (light shaded) 
  confidence allowed regions for $\Omega_m$ and $w_Q$ using simulated
  data from the SNAP and SNfactory observatories, with lensing effects
  from 90\% NFW halos and 10\% point masses included. The simulated
  data is generated with $\Omega_m = 0.3$ and $w_Q = -1$.}
\label{fig7}
\end{figure}

\acknowledgments

We wish to thank A.~Melchiorri for valuable discussions during the
initial stages of the project.

\newcommand\AJ[3]{~Astron. J.{\bf ~#1}, #2~(#3)}
\newcommand\APJ[3]{~Astrophys. J.{\bf ~#1}, #2~ (#3)}
\newcommand\apjl[3]{~Astrophys. J. Lett. {\bf ~#1}, L#2~(#3)}
\newcommand\ass[3]{~Astrophys. Space Sci.{\bf ~#1}, #2~(#3)}
\newcommand\cqg[3]{~Class. Quant. Grav.{\bf ~#1}, #2~(#3)}
\newcommand\mnras[3]{~Mon. Not. R. Astron. Soc.{\bf ~#1}, #2~(#3)}
\newcommand\mpla[3]{~Mod. Phys. Lett. A{\bf ~#1}, #2~(#3)}
\newcommand\npb[3]{~Nucl. Phys. B{\bf ~#1}, #2~(#3)}
\newcommand\plb[3]{~Phys. Lett. B{\bf ~#1}, #2~(#3)}
\newcommand\pr[3]{~Phys. Rev.{\bf ~#1}, #2~(#3)}
\newcommand\PRL[3]{~Phys. Rev. Lett.{\bf ~#1}, #2~(#3)}
\newcommand\PRD[3]{~Phys. Rev. D{\bf ~#1}, #2~(#3)}
\newcommand\prog[3]{~Prog. Theor. Phys.{\bf ~#1}, #2~(#3)}
\newcommand\RMP[3]{~Rev. Mod. Phys.{\bf ~#1}, #2~(#3)}

\end{document}